\documentclass[%
reprint,
showpacs,
showkeys,
onecolumn,
amsmath,amssymb,
aps,
]{revtex4-1}
\usepackage{graphicx}\usepackage{color}
\usepackage{amsmath}
\usepackage{mathtools}
\usepackage[bottom]{footmisc}

\def\bea{\begin{eqnarray}}
\def\eea{\end{eqnarray}}
\begin{document}
%\documentclass[ paper,onecolumn,12pt, nofootinbib]{revtex4-1}
%\usepackage{graphicx}\usepackage{color}

%\def\bea{\begin{eqnarray}}
%\def\eea{\end{eqnarray}}
%\begin{document}

\title{Non-linear interaction of laser light with vacuum: contributions to the energy density and pressure in presence of an intense magnetic field}

\author{ M. A. P\'erez-Garc\'ia$^1$, A. Pérez Martínez \footnotetext{On leave from Departamento de Fisica Teorica,Instituto de Cibernetica Matematica y Fisica (ICIMAF), Calle E esq 15 No. 309 Vedado, La Habana, 10400, Cuba} $^{1a}$, E. Rodríguez Querts$^2$}

\affiliation{$^1$Department of Fundamental Physics and IUFFyM, University of Salamanca, 
Plaza de la Merced s/n, E-37008 Salamanca, Spain\\$^2$Departamento de Fisica Teorica,\\ Instituto de Cibernetica Matematica y Fisica (ICIMAF), \\ Calle E esq 15 No. 309 Vedado, La Habana, 10400, Cuba}

\date{\today}

\begin{abstract}

Recent simulations show that very large electric and magnetic fields near the kilo Tesla strength will likely be generated by ultra-intense lasers at existing facilities over distances of hundreds of microns in underdense plasmas. Stronger ones are even expected in the future although some technical dificulties must be overcome. In addition, it has been shown that vacuum exhibits a peculiar non-linear behaviour in presence of high magnetic and electric field strengths. In this work we are interested in the analysis of thermodynamical contributions of vacuum to the energy density and pressure when radiation interacts with it in the presence of an external magnetic field. Using the Euler-Heisenberg formalism in the regime of weak fields i.e. smaller than critical Quantum Electrodynamics field strength values, we evaluate  these magnitudes and analyze the highly anisotropic behaviour we find. Our work has implications for photon-photon scattering with lasers and astrophysically magnetized underdense systems far outside their surface where matter effects are increasingly negligible.
\end{abstract}

\maketitle

\section{Introduction}

%%%%%%%%%%%%%%

The ultra-high fields of high-power short-pulse lasers are expected to contribute to the understanding of fundamental properties of the quantum vacuum and quantum theory in the presence of very strong fields. Such large magnetic and electric field values can be obtained on Earth laboratories using modern laser facilities \cite{tomma}. Quoted fields at the kiloTesla field strength have been simulated from the strong coupling of laser angular momentum to plasma through ponderomotive forces, resulting in axial magnetic fields sustained in $\sim$200 $\mu$m in length, and lasting several picoseconds \cite{andrew}. However, they remain lower than the predicted values of critical field strengths $B_q\sim 4.41\times 10^{13}$ $G$ and associated $E_q\sim 1.3 \times 10^{18}$ $V/m$ where a virtual  electron-positron $e^+e^-$ pair gains its rest mass energy over a Compton wavelength,  $\lambda_e=\hbar/m_e c$, being $m_e$ the electron mass, and materializes as a real pair. In nature these fields have only been measured appearing  in astrophysical systems such as white dwarfs or neutron stars \cite{ns}. 
For example, when a neutron star collapses, its angular momentum and its magnetic field are transferred onto the resulting star, and thereby the magnetic flux density increases dramatically, due to the smaller size of the stellar remnant \cite{ns1}. Values beyond $\sim 10^2 B_q$ i.e. around $B\sim 10^{15}$ G  have been reported in magnetars. Inside those objects matter is thought to be partially magnetized \cite{perez}.  In such scenarios the electromagnetic field has a long-scale variation and changes over distances much greater than the quoted electron Compton wavelength,  so for all purposes it can be considered to be homogeneous and constant with time. 

It is worth mentioning that most processes outside these dense magnetized stars are dominated by their huge magnetic field. The induced electrostatic forces on an electron near the surface are many orders of magnitude larger than the gravitational forces and lead to an extraction of particles from the stellar surface, which then fill the surrounding magnetosphere. The actual configuration of the (multipolar) field is complex but early calculations by Deutsch \cite{deu} showed  that a sufficiently magnetized rotating star will not continue to stay in vacuum, but surround itself with a plasma charge cloud. Low density plasma will fill the space and will affect the radiation as it travels to eventually reach us on Earth. As shown in the literature \cite{Elizabeth1}\cite{Villalba-Chavez:2010jvj}, it is expected that directions parallel and perpendicular to the magnetic field lines will affect differently to the propagation of electromagnetic radiation constituting the so-called birefringence. Despite many efforts its experimental direct detection remains yet pending, see a discussion in \cite{bire1}.  When considering these huge values of fields near or above critical values one must pay special attention to the quantum effects from vacuum structure \cite{Chaichian}.  In this way the theory of Quantum Electrodynamics \cite{qed} (QED) or Born-Infeld theory \cite{bi1,bi2}  have proven to be a useful tool when attempting to describe phenomenology such as in photon-photon scattering scenarios. However, generally speaking, effects from  the vacuum in non-linear electrodynamics experiments are extremely small and the detection of these are a real challenge.

One of the key quantities regarding the impact of non-linear behaviour is the vacuum polarization. This radiative correction modifies the propagation of photons in vacuum through the appearance of virtual loops of fermion anti-fermion pairs (typically electron-positron pairs). These corrections will induce a charge renormalization due to polarization screening that will ultimately translate into a distance dependent value of the fine structure constant $\alpha(r)$ with $r$ being the length from the {\it dressed} particle. Here the term dressed refers to the actual particle plus the cloud of virtual particles around it as it emerges in a many-body theory in the concept of {\it quasi-particles}. When $r\ll \lambda_e$ then $\alpha\sim 1/137$ however for smaller distances at the weak boson $Z_0$ energy scale $\alpha (\lambda_Z)>\alpha$. On the terrestrial side, ultra-intense fields generated with lasers form a virtual cloud of particles. A world-wide class of rising intensity lasers include the Exawatt ($10^{18}$ W) lasers belonging to  the ELI beamlines in Europe \cite{eli}  or those in LFEX Petawatt facility in Japan \cite{lfex}. They are expected to approach the Schwinger limit corresponding to an intensity $I\sim 10^{33} \,W/m^2$ and associated critical field strengths in the future. Namely this is possible due to the  Chirped Pulse Amplification (CPA) proposed by G. Mourou in 1985 \cite{mouru}. The highest intensity presently reached is about $\sim 2$ PW  with the LFEX Petawatt facility.

A way to probe the structure of quantum vacuum is coupling the polarization insertion with additional external photons.  Then a plethora of non-linear effects should appear from photon-photon scattering. In the very early times this was described using the Euler-Heisenberg approach \cite{eh} using field invariants, $F, G$ as will be described later. Note this formulation does not provide full account for all quantum effects at energies comparable or larger than electron's rest mass $m_ec^2$ and stronger fields, close to the critical values, where  a more refined approach is needed \cite{bire1}. 
 
In this work we will focus on yet another manifestation of the vacuum anisotropic behaviour \cite{rojas}, as we will be interested in obtaining the contribution of the non-linear vacuum to the thermodynamical properties of pressure, $P$, and energy density, $\rho$,  in the presence of a subcritical external magnetic field as radiation fills the space. We will discuss how sizable this contribution is and what are the laser intensities needed in  laboratories to {\it indirectly} check the effects leading to birefringence of vacuum as a first step towards a more elaborated  experimental  determination.

\section{Formalism}

We will be considering the Euler-Heisenberg (EH) formalism that we summarize in what follows for the sake of completeness.  The lagrangian model can be cast into a functional dependent form by using two electromagnetic invariants $F=-\frac{1}{4} F^{\mu \nu}F_{\mu \nu}=\epsilon_{0} E^{2}-\frac{B^{2}}{\mu_{0}}$ and $G=-\frac{1}{4} F^{\mu \nu} \tilde{F}_{\mu \nu}=\sqrt{\frac{\epsilon_{0}}{\mu_{0}}}(\boldsymbol{E} \cdot \boldsymbol{B})$ where $F^{\mu \nu}=\partial^{\mu} A^{\nu}-\partial^{\nu} A^{\mu}$ is the electromagnetic field strength tensor written in terms of fluctuating fields  $A^{\mu}$ and $\tilde{F}^{\mu \nu}=\epsilon^{\mu \nu \alpha \beta} F_{\alpha \beta} / 2$ is the dual tensor. Note that other works may use a different convention when defining $F,G$.
In our approach we will be considering that the field $A^{\mu}=A^{\mu}_e+A_{w}^{\mu}$, is a composition of a classical background laser field,  $A^{\mu}_e$, and the probing photon field, $A_{w}^{\mu}$. Therefore we will have in terms of Feynman diagrams a leading order type with four photon scattering amplitudes replaced by only photon-background (laser) field scattering and there is no need to resolve fermion loops as we will restrict to energies not capable of creating electron-positron pairs.

Keeping the quadratic order terms in $\mathrm{F}$ and $\mathrm{G}$ we can write the EH lagrangian valid in the weak field limit $\left(B \ll B_{q}, E \ll E_{q}\right)$ as,
\begin{equation}
\mathcal{L}_{E H}=\mathcal{L}_{0}+\xi_L \mathcal{L}_{0}^{2}+ \xi_T \frac{7  \ \epsilon_{0}^{2} c^{2}}{4}(\boldsymbol{E} \cdot \boldsymbol{B})^{2},
\end{equation}

with $\mathcal{L}_{0}=\frac{\epsilon_{0}}{2}\left(E^{2}-c^{2} B^{2}\right)$ the classical Lagrangian density and $\epsilon_{0}$ and $c$ are the dielectric constant and the speed of light in the vacuum, respectively. In addition $c^2=1/\epsilon_0 \mu_0$, being $\mu_0$ the vacuum permeability. Vacuum corrections emerge through the parameters, $\xi_L,\xi_T$ that we will take as equal in our approach $\xi_L=\xi_T=\xi$  and $\xi=\frac{8 \alpha^{2} \hbar^{3}}{45 m_{e}^{4} c^{5}} \sim 6.7 \times 10^{-30} \frac{m^{3}}{J}$.

The propagation of a probe photon (wave) in vacuum in presence of an external electromagnetic field is not trivial as it will depend on the actual strength and orientation of the field. In our work and for the sake of simplicity we will assume that there are fluctuating electric and magnetic fields, $\boldsymbol{E}_{w}$ and $\boldsymbol{B}_{w}$, and an external magnetic field denoted by $\boldsymbol{B}_{e}$. So we have that the sum values are  $\boldsymbol{E}_t=\boldsymbol{E}_{w}$ and $\boldsymbol{B}_{t}=\boldsymbol{B}_{w}+\boldsymbol{B}_{e}$. Photons propagating in vacuum parallel to the magnetic field $\left(\boldsymbol{k} \| \boldsymbol{B}_{e}\right)$ do not feel the effect of the field and move freely obeying the light cone dispersion equation with velocity $c$, while photon propagation perpendicular to the magnetic field does imply changes in dispersion velocity. The latter is yet another  manifestation of the vacuum birefringence, describing the fact  that photons can be polarized in two modes determined by the orientation of the external and wave magnetic fields, respectively.  We label mode perpendicular ($\perp$) when the $\boldsymbol{E_{w}} \| \boldsymbol{B_{e}}$ along with  $\boldsymbol{B_{w}} \perp \boldsymbol{B_{e}}$ and mode parallel ($\|$) when $\boldsymbol{E_{w}} \perp \boldsymbol{B_{e}}$ and $\boldsymbol{B_{w}} \| \boldsymbol{B_{e}}$. Without loss of generality we choose using an external field in the Cartesian $z-$direction $\boldsymbol{B}_e={B}_e \boldsymbol{k}$.

The energy momentum tensor (EMT)  $T^{\mu \nu}$ is obtained from the variation of the action functional derived from the  EH effective lagrangian $\mathcal{L}_{E H}$. In our convention we use a Minkowski metric tensor  $g^{\mu \nu}=\operatorname{diag}(+1,-1,-1,-1)$. Then the EMT reads as
\begin{equation}
  \begin{aligned}
T^{\sigma \delta} =-\left[ F_{\beta}^{ ~\sigma}F^{\beta \delta}-\frac{1}{4}g^{\sigma \delta} F_{\mu \nu}F^{\mu \nu}\right]+
\xi [ -\frac{1}{64} (-4 g^{\sigma \delta} F_{\mu \nu}F^{\mu \nu}-32 F_{\mu}^{ ~\sigma}F^{\delta \mu})F_{\beta \lambda}F^{\beta \lambda}+\\
 \frac{7}{64}(-g^{\sigma \delta} \tilde{F}_{\mu \nu}F^{\mu \nu}-2\tilde{F}_{\mu}^{ ~\sigma}F^{\delta \mu}+6 F^{\sigma \mu} \tilde{F}_{\mu}^{ ~\delta}  ) \tilde{F}_{\beta \nu}F^{\beta \nu} ],
\end{aligned}
\end{equation}
%\end{equation}
where indexes $\mu, \nu=0, 1,2,3$. From the diagonal terms of the stress-energy momentum tensor  we obtain the energy density $\rho$ and pressure $P$ as follows
\begin{equation}
\rho=T^{00}, \quad P=\frac{1}{3} \sum_{k=1}^{3} T^{kk}.
\end{equation}

Note that for each of the latter magnitudes we have subsequent modes $\rho_\perp,\rho_\|$ and  $P_\perp,P_\|$ due to the fixed orientation of an external magnetic field in the $z$-axis direction. Therefore, in the weak field limit the components of the EMT can be split up into perpendicular and parallel ones by considering the relative orientation of $\boldsymbol{B_e,B_w,E_e}$ described before. As a consequence it follows that up to $F^2, G^2$ order (we set $c=\epsilon_0=\mu_0=1$ in what follows) the general 
form is the contribution of a classical linear term (setting $\xi=0$) plus that of non-linear (NL) terms in the EH lagrangian. Explicitly

\begin{equation}
\rho=T^{00}=T_{\xi=0}^{00}+\xi T_{NL}^{00},
\end{equation}

\begin{equation}
 P=\frac{1}{3} \sum_{k=1}^{3} T_{\xi=0}^{kk}+\xi T_{NL}^{kk}.
\end{equation}

In particular we find

  \begin{equation}
  \begin{aligned}
%\begin{equation}
T_{\|}^{00}=\frac{1}{2} B_{e}^{2}+B_{e} B_{w} +\frac{1}{2} B_{w}^{2}+\frac{1}{2} E_{w}^{2} +\xi (-\frac{1}{2} B_{e}^{2} E_{w}^{2} -B_{e} B_{w} E_{w}^{2} -\frac{1}{2} B_{w}^{2} E_{w}^{2} +\frac{3}{4} E_{w}^{4} -\frac{1}{4} B_{e}^{4} ),
%B_{e}^{3} B_{w} \xi -\frac{3}{2} B_{e}^{2} B_{w}^{2} \xi -B_{e} B_{w}^{3} \xi -\frac{1}{4} B_{w}^{4}
  \end{aligned}
\end{equation}

\begin{equation}
T_{\bot}^{00}=\frac{1}{2} B_{e}^{2}+\frac{1}{2} B_{w}^{2}+\frac{1}{2} E_{w}^{2} +\xi(-\frac{1}{4} B_{e}^{4} -\frac{1}{2} B_{e}^{2} B_{w}^{2} +\frac{5}{4} B_{e}^{2} E_{w}^{2}-\frac{1}{4} B_{w}^{4} -\frac{1}{2} B_{w}^{2} E_{w}^{2} +\frac{3}{4} E_{w}^{4}),
\end{equation}

  \begin{equation}
  \begin{aligned}
%\begin{equation}
T_{\|}^{11}=\frac{1}{2} B_{e}^{2}+B_{e} B_{w} +\frac{1}{2} B_{w}^{2}-\frac{1}{2} E_{w}^{2}+\xi(-\frac{3}{4} B_{e}^{4}  -3 B_{e}^{3} B_{w} -\frac{9}{2} B_{e}^{2} B_{w}^{2} +\frac{3}{2} B_{e}^{2} E_{w}^{2} -3 B_{e} B_{w}^{3}\\+3 B_{e} B_{w} E_{w}^{2} -\frac{3}{4} B_{w}^{4}+\frac{3}{2} B_{w}^{2} E_{w}^{2} -\frac{3}{4} E_{w}^{4}) ,
  \end{aligned}
\end{equation}

\begin{equation}
T_{\bot}^{11}=\frac{1}{2} B_{e}^{2}-\frac{1}{2} B_{w}^{2}+\frac{1}{2} E_{w}^{2}+ \xi (-\frac{3}{4} B_{e}^{4}-\frac{1}{2} B_{e}^{2} B_{w}^{2} -\frac{5}{4} B_{e}^{2} E_{w}^{2}+\frac{1}{4} B_{w}^{4} -\frac{1}{2} B_{w}^{2} E_{w}^{2} +\frac{1}{4} E_{w}^{4}),
\end{equation}

  \begin{equation}
  \begin{aligned}
T_{\|}^{22}=\frac{1}{2} B_{e}^{2}+B_{e} B_{w} +\frac{1}{2} B_{w}^{2}+\frac{1}{2} E_{w}^{2}+\xi( -\frac{3}{4} B_{e}^{4} -3 B_{e}^{3} B_{w} -\frac{9}{2} B_{e}^{2} B_{w}^{2} +\frac{1}{2} B_{e}^{2} E_{w}^{2} -3 B_{e} B_{w}^{3}\\+B_{e} B_{w} E_{w}^{2} -\frac{3}{4} B_{w}^{4} +\frac{1}{2} B_{w}^{2} E_{w}^{2} +\frac{1}{4} E_{w}^{4} ),
  \end{aligned}
\end{equation}

\begin{equation}
T_{\bot}^{22}=\frac{1}{2} B_{e}^{2}+\frac{1}{2} B_{w}^{2}+\frac{1}{2} E_{w}^{2}+\xi(-\frac{3}{4} B_{e}^{4} -\frac{3}{2} B_{e}^{2} B_{w}^{2} -\frac{5}{4} B_{e}^{2} E_{w}^{2} -\frac{3}{4} B_{w}^{4}+\frac{1}{2} B_{w}^{2} E_{w}^{2} +\frac{1}{4} E_{w}^{4}),
\end{equation}

  \begin{equation}
  \begin{aligned}
T_{\|}^{33}=  -\frac{1}{2} B_{e}^{2}-B_{e} B_{w} -\frac{1}{2} B_{w}^{2}+\frac{1}{2} E_{w}^{2} +\xi( \frac{1}{4} B_{e}^{4}  +B_{e}^{3} B_{w} +\frac{3}{2} B_{e}^{2} B_{w}^{2} -\frac{1}{2} B_{e}^{2} E_{w}^{2} +B_{e} B_{w}^{3}\\ -B_{e} B_{w} E_{w}^{2} +
\frac{1}{4} B_{w}^{4} -\frac{1}{2} B_{w}^{2} E_{w}^{2} +\frac{1}{4} E_{w}^{4} ),
  \end{aligned}
\end{equation}

\begin{equation}
T_{\bot}^{33}= -\frac{1}{2} B_{e}^{2}+\frac{1}{2} B_{w}^{2}-\frac{1}{2} E_{w}^{2}+\xi( \frac{1}{4} B_{e}^{4} -\frac{1}{2} B_{e}^{2} B_{w}^{2} -\frac{5}{4} B_{e}^{2} E_{w}^{2} -\frac{3}{4} B_{w}^{4} +\frac{3}{2} B_{w}^{2} E_{w}^{2} -\frac{3}{4} E_{w}^{4} ).
\end{equation}

It is worth to emphasize at this point that the non-linear effects fully depend on the $\xi$ parameter strength  being thus tiny corrections even for fields close to the  critical  values in the scenario we describe. As an example, using the electric field $E_{w}$ associated to the wave, $E_w\sim B_{w}=\sqrt{\frac{I}{2}}$ we find $\Delta T_{\|}^{00}=T_{\|}^{00} -T_{\xi=0}^{00} \sim 10^{-8} T_{\xi=0}^{00}$, $\Delta T_{\bot} \sim 10^{-5}T_{\xi=0}^{00}$, $\Delta P_{\|}\sim 10^{-8}P_{\xi=0}$, $\Delta P_{\bot}\sim 10^{-5}P_{\xi=0}$ for an upper limiting critical value $I\sim 10^{33}$ $W/m^2$ and $B_e=100$ T. Thus  it seems to be very challenging to experimentally determine in the laboratory where presently achieved intensities are many orders of magnitude lower, e.g. around $\sim 10^{27}$ $W/m^2$ in the ELI Beam lines.

\section{Results}
In this section we analyze the results and particularize imposing  that the probe photons carry an associated electromagnetic plane wave, whose fields $\boldsymbol{E}_{w}$ and  $\boldsymbol{B}_{w}$ fulfill the relation $\epsilon_{0} E_{w}-\frac{B_{w}}{\mu_{0}}=0$.  We will mostly  consider scenarios with external laser fields $B_e < 12$ T being homogeneous and static as time variation is much smaller than their typical inverse frequencies $\sim 1/\nu$.  However, as mentioned,  fields $B_e\sim 10^2-10^3$ T are likely possible in the near future. We will remain in the weak field regime with intensities  $I <10^{22}\,W/m^2$.

%%%%%%%%%%%%%%%%%%%%%%%%%%%%%%%%%%%%%%%%%%%%%%%%%%%%%%%%%%%%%%%%
\begin{figure}[h!]
\begin{center} 
\includegraphics [angle=0,scale=0.65]{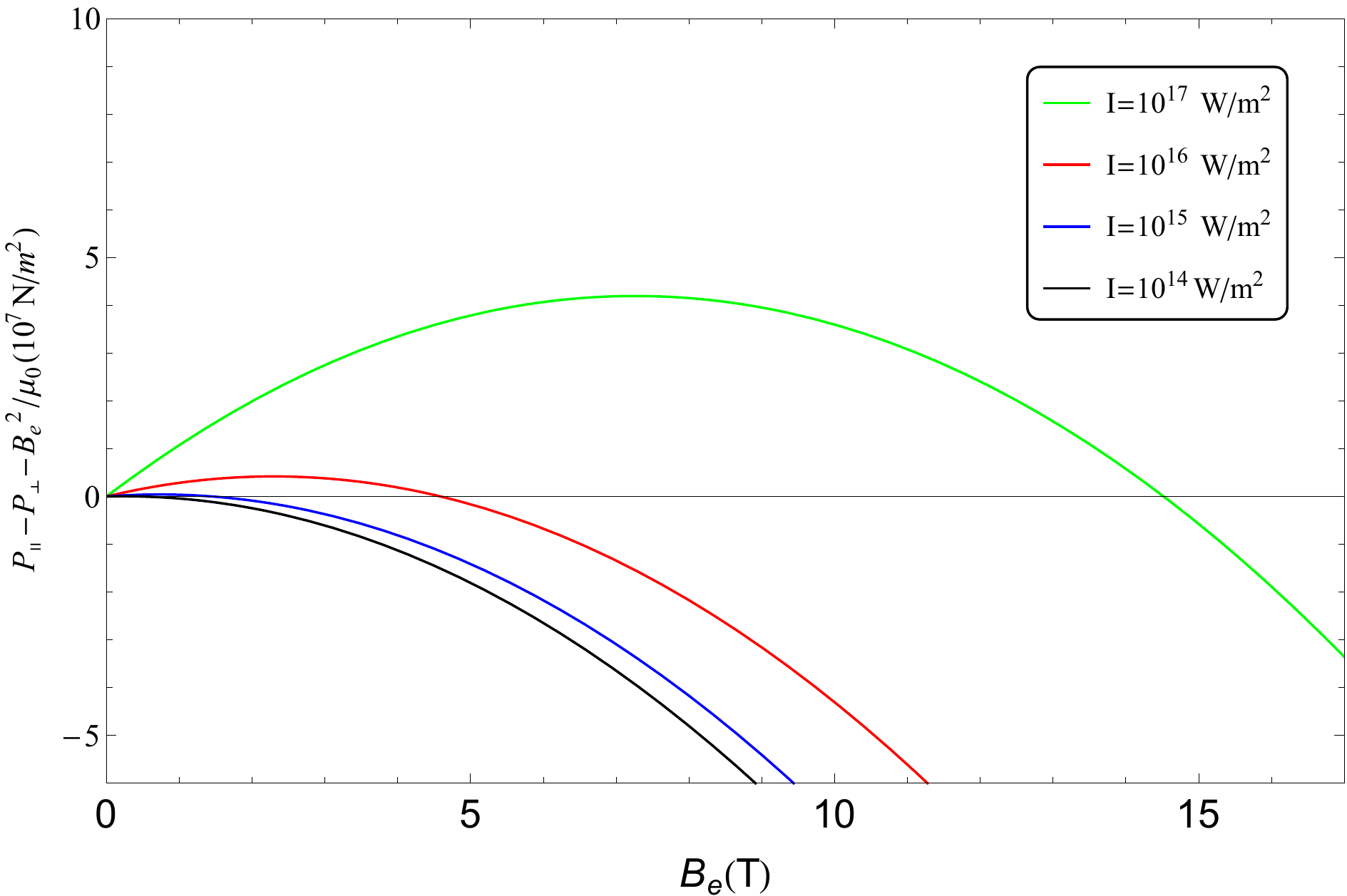}%{fig1a.eps}
\caption{Difference of parallel and perpendicular pressures, $P_{\|}-(P_\perp +B_e^2/\mu_0)$,  in units of $10^{7}  \,N/m^2$ as a function of external magnetic field  strength $B_e$, in T, for intensities $I=10^{14}, 10^{15}, 10^{16}, 10^{17}$ $\rm W/m^2$. See text for details.}
\label{Fig1}
\end{center}
\end{figure}
%%%%%%%%%%%%%%%%%%%%%%%%%%%%%%%%
We show in Fig. \ref{Fig1} the difference $P_{\|}-(P_\perp +B_e^2/\mu_0)$ in units of $10^{7} \,N/m^2$ versus the external field $B_e$ in T. We select different laser intensities $I=10^{14}, 10^{15}, 10^{16}, 10^{17}$ $W/m^2$. We can observe that there is a change in sign (from positive to negative values)  for a given intensity $I$ as external magnetic field strength grows. For $I=10^{14}$ $W/m^2$ it is hardly visible on the plot but the crossing value is $B_e=0.33$ T. This behaviour signals the conditions of instability among the parallel and perpendicular components in pressure. Note that an additional contribution from $B_e^2/\mu_0$ has been added in order to compare to the so-called parallel firehose instability \cite{firehose} in weakly collisional plasma flows but in this case restricting only to contributions from vacuum. 

%%%%%%%%%%%%%%%%%%%%%%%%%%%%%%%%%%%%%%%%%%%%%%%%%%%%%%%%%%%%%%%%
\begin{figure}[h!]
\begin{center} 
\includegraphics [angle=0,scale=0.65]{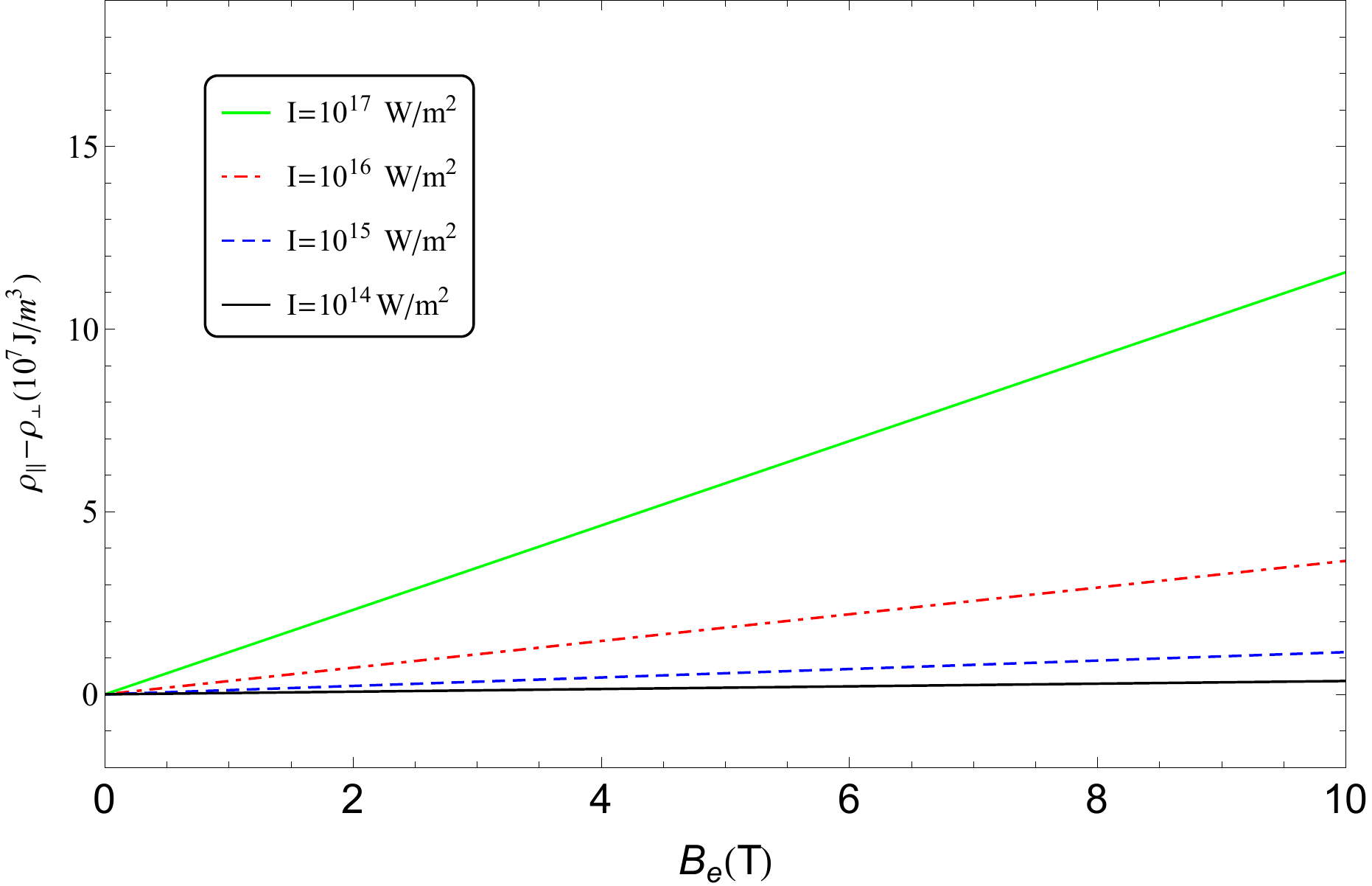}%{fig2a.eps}
\caption{Difference of energy density contribution from vacuum for  parallel and perpendicular components in units of $10^{7} \,J/m^3$ as a function of external $B_e$ field in T for intensities $I=10^{14}, 10^{15}, 10^{16}, 10^{17}$ $W/m^2$. See text for details.}
\label{Fig2}
\end{center}
\end{figure}
%%%%%%%%%%%%%%%%%%%%%%%%%%%%%%%%
We show in Fig. \ref{Fig2} the difference of vacuum contributions to the parallel and perpendicular energy density $\rho$ (or $T_{00}$)  in units of $10^{7} \,J/m^3$ versus the external field $B_e$ in T. We select different laser intensities $I=10^{14}, 10^{15}, 10^{16}, 10^{17}$ $W/m^2$. We can observe that the parallel component is always larger than the perpendicular one. As expected, for the lower intensity  and lower $B_e$ strength cases considered we obtain  that the two components tend to become similar $\rho_\perp\sim \rho_\|$ decreasing the impact of anisotropy.

%%%%%%%%%%%%%%%%%%%%%%%%%%%%%%%%%%%%%%%%%%%%%%%%%%%%%%%%%%%%%%%%
\begin{figure}[h!]
\begin{center} 
\includegraphics [angle=0,scale=0.65]{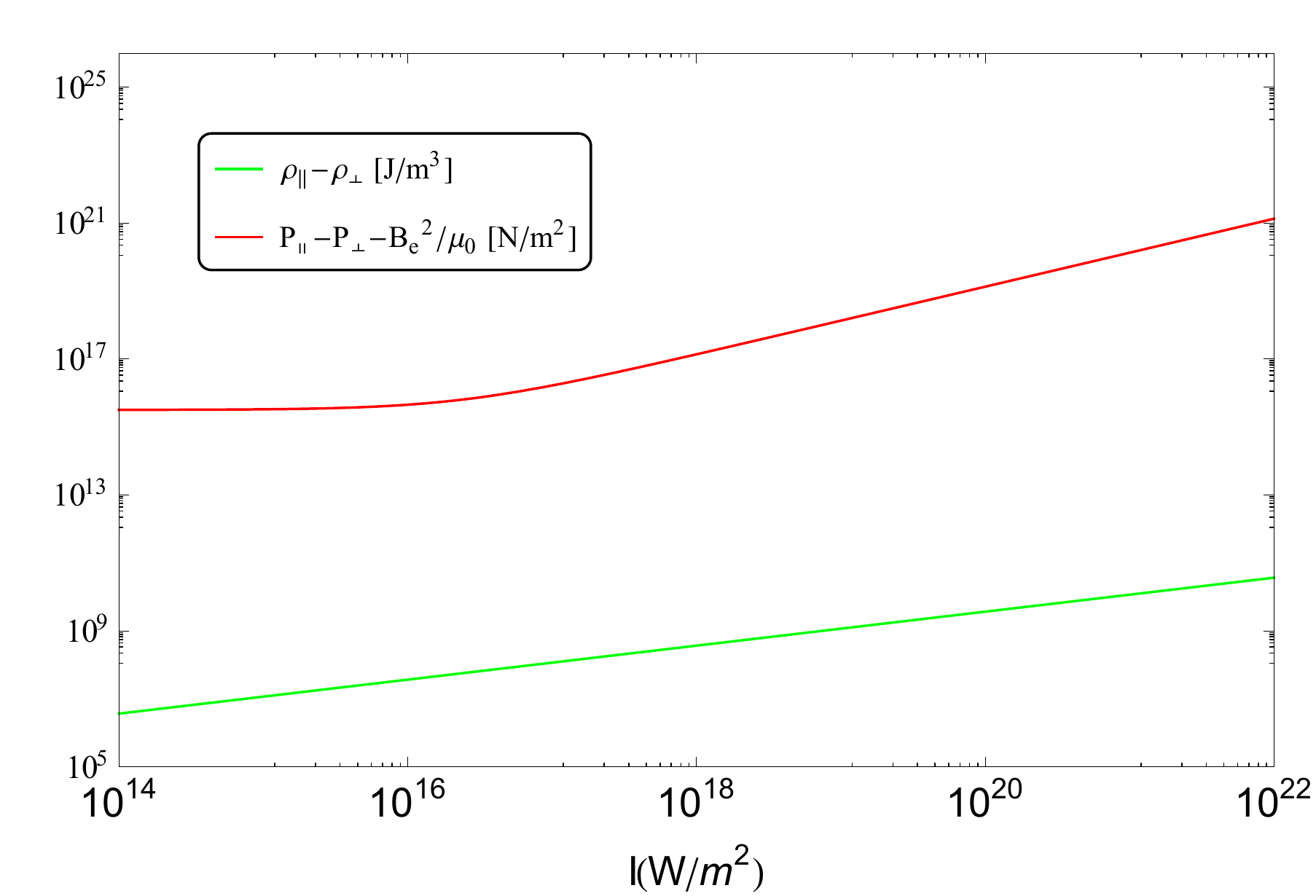}%{fig3a.eps}
\caption{Value of pressure difference and energy density difference contributions in units of $N/m^2 \,(J/m^3)$ from vacuum as a function of the logarithm (base 10) of  the intensity $I$ in $W/m^2$. $B_e=10$ T.}
\label{Fig3}
\end{center}
\end{figure}
%%%%%%%%%%%%%%%%%%%%%%%%%%%%%%%%
We show in Fig. \ref{Fig3} the  vacuum contributions to the value of $P_{\|}-(P_\perp +B_e^2/\mu_0)$ and difference of  perpendicular and parallel components,  in $\rho_\|-\rho_{\bot}$ in units of $\,N/m^2 \,(J/m^3)$, respectively, as a function of the logarithm (base 10) of  the intensity $I$ in $W/m^2$. We set an external field value $B_e=10$ T.

It is important to note at this point that the non-linear effects are parameterized in the $\xi$ value. Although fixed equal in QED, additional freedom in the choice of $\xi_L$ and $\xi_T$ could actually yield an independent test of individual lagrangian terms and non-linear effects. We conclude that these vacuum contributions will be challenging task for the future facilities but will provide valuable help in the understanding on the role of non-linear effects on the anisotropy of vacuum to external photon probes.

%%%%%%%%%%%%%%%%%%%

\section*{Acknowledgments}

MAPG and APM  would like to acknowledge financial support from Junta de Castilla y Le\'on through the grant SA096P20, Agencia Estatal de Investigaci\'on through the grant PID2019-107778GB-100, Spanish Consolider MultiDark FPA2017-90566-REDC and PHAROS COST Actions MP1304 and CA16214.

%###############################################################

\end{document}